\documentclass[10pt, onecolumn]{IEEEtran}
\usepackage{qtree}
\usepackage{bbm}
\usepackage{amsmath}
\usepackage{amssymb}
\usepackage{amsfonts}
\usepackage{epsfig}
\usepackage{calc}
\usepackage{color}
\usepackage[all]{xy}
\usepackage{bm}
\usepackage{algorithmicx}
\usepackage{algpseudocode}
\usepackage{algorithm}
\usepackage{enumerate}
\usepackage{pdfsync}
\usepackage{framed}
\usepackage{caption}
\usepackage{amsthm}
\usepackage{blkarray}
\usepackage{subcaption}
\usepackage{enumitem}
\usepackage{hyperref}
\usepackage{mathrsfs}
\usepackage{caption}
\usepackage{float}
\newtheorem{defn}{Definition}

\newtheorem{thm}{Theorem}[section]
\newtheorem{cor}[thm]{Corollary}
\newtheorem{prop}{Proposition}

\newtheorem{lem}[thm]{Lemma}
\newtheorem{conj}[thm]{Conjecture}
\newtheorem{constr}[thm]{Construction}
\usepackage[normalem]{ulem}
\newtheorem{proposition}[thm]{Proposition}
\newtheorem{remark}{Remark}[section]
\newtheorem{example}{Example}[section]

\newcounter{definition}[section]
\newenvironment{definition}[1][]{\refstepcounter{definition}\par\medskip
   \noindent \textbf{Definition~\thesection.\thedefinition. #1} \rmfamily}{\medskip}

\newcommand{\bit}{\begin{itemize}}
\newcommand{\eit}{\end{itemize}}
\newcommand{\bcor}{\begin{cor}}
\newcommand{\ecor}{\end{cor}}
\newcommand{\beq}{\begin{equation}}
\newcommand{\eeq}{\end{equation}}
\newcommand{\beqn}{\begin{equation*}}
\newcommand{\eeqn}{\end{equation*}}
\newcommand{\bea}{\begin{eqnarray}}
\newcommand{\eea}{\end{eqnarray}}
\newcommand{\bean}{\begin{eqnarray*}}
\newcommand{\eean}{\end{eqnarray*}}
\newcommand{\ben}{\begin{enumerate}}
\newcommand{\een}{\end{enumerate}}
\newcommand{\bdefn}{\begin{defn}}
\newcommand{\edefn}{\end{defn}}
\newcommand{\bnote}{\begin{remark}}
\newcommand{\enote}{\end{remark}}
\newcommand{\bprop}{\begin{prop}}
\newcommand{\eprop}{\end{prop}}
\newcommand{\blem}{\begin{lem}}
\newcommand{\elem}{\end{lem}}
\newcommand{\bthm}{\begin{thm}}
\newcommand{\ethm}{\end{thm}}
\newcommand{\bconj}{\begin{conj}}
\newcommand{\econj}{\end{conj}}
\newcommand{\bconstr}{\begin{constr}}
\newcommand{\econstr}{\end{constr}}
\newcommand{\bpf}{\begin{proof}}
\newcommand{\epf}{\end{proof}}

\title{A Study on the Impact of Locality in the Decoding of Binary Cyclic Codes}

\begin{document}

\author{
	\IEEEauthorblockN{M. Nikhil Krishnan$^a$, Bhagyashree Puranik$^b$, P. Vijay Kumar$^c$, Itzhak Tamo$^d$, and Alexander Barg$^e$}
	\thanks{$^a$ Department of Electrical Communication Engineering, Indian Institute of Science, Bangalore, India, Email: nikhilkrishnan.m@gmail.com. Research supported through Visvesvaraya PhD Scheme for Electronics \& IT awarded by Department of Electronics and Information Technology, Govt. of India.
	
	$^b$  Department of Electrical Communication Engineering, Indian Institute of Science, Bangalore, India, Email: bhagya.puranik@gmail.com
	
	$^c$ Department of Electrical Communication Engineering, Indian Institute of Science, Bangalore, India, Email: pvk1729@gmail.com. P. V. Kumar is also an Adjunct Research Professor at the University of Southern California.  His research is supported in part by NSF under Grant No. 1421848 and in part by the joint UGC-ISF research program.
	
	$^d$ Department of EE-Systems, Tel Aviv University, Tel Aviv, Israel, Email: zactamo@gmail.com. Research is supported by ISF grant no.~1030/15 and the NSF-BSF grant no.~2015814.
	
	$^e$ Dept. of ECE and ISR, University of Maryland, College Park, USA, and Inst. Inform. Trans. Probl. (IITP), Russian Academy of Sciences, Moscow, Russia, Email: abarg@umd.edu. Research is supported in part by NSF grants CCF1422955 and CCF1618603.}
%
%
%
%
}	

	\maketitle

\begin{abstract}
	In this paper, we study the impact of locality on the decoding of binary cyclic codes under two approaches, namely ordered statistics decoding (OSD) and trellis decoding. Given a binary cyclic code having locality or availability, we suitably modify the OSD to obtain gains in terms of the Signal-To-Noise ratio, for a given reliability and essentially the same level of decoder complexity.  With regard to trellis decoding, we show that careful introduction of locality results in the creation of cyclic subcodes having lower maximum state complexity. We also present a simple upper-bounding technique on the state complexity profile, based on the zeros of the code.  Finally, it is shown how the decoding speed can be significantly increased in the presence of locality, in the moderate-to-high SNR regime, by making use of a quick-look decoder that often returns the ML codeword.
 
\end{abstract}

\section{Introduction}\label{sec:intro}
Efficient decoding of general linear block codes has historically been a well-studied problem. {\color{black} The decoding procedure
that optimizes the performance of the code is maximum likelihood (ML) decoding. Even though optimal, ML decoding schemes in general become computationally complex as the code-length increases. One of the most developed research directions
aimed at reducing the complexity of ML decoding has been trellis decoding and associated trellis representations of codes 
\cite{BCJR,Wolf}. Several schemes that approximate ML decoding have been suggested in the literature; among them such reliability-based  decoding methods as the Chase algorithm \cite{Chase72}, decoding based on ordered statistics \cite{FosLin}, and others \cite[Ch.~10]{LinCos}. }

At the same time, significant progress has been achieved in decoding particular code families based on the constraints for the 
parity check matrix or other structural properties of the codes. Arguably the most far-reaching results along these lines have been accomplished for 
low-density-parity-check (LDPC) codes and related families based on their iterative decoding \cite{HagOffPap}. Iterative decoding
schemes have also been applied to other codes and a variety of other problems in information and communication theory.

The focus of this paper is a class of codes introduced recently by Gopalan et al. \cite{GopHuaSimYek} for applications in distributed storage. The main idea of \cite{GopHuaSimYek} stems from the need to minimize network communication between storage nodes and
is formalized in the concept of locality wherein an erased code-symbol can be repaired using a small subset of other symbols of the codeword, usually called the recovery set of the symbol. In the context of linear codes, locality translates into the requirement that every coordinate of the codeword
appear in a low-weight parity check equation of the code. Thus, while we do not require that all the parities are of low weight, 
a linear locally recoverable code (linear LRC code) has a set of low-weight parities whose union contains all the code coordinates. 
This feature bridges the classes of LRC codes and LDPC codes, and has been used in previous research \cite{Agarwal2016,TBF16} to prove
bounds on the rate of LRC codes with a given minimum distance.

The described connection between LRC and LDPC codes is the starting point of this research which develops in the context
of Ordered Statistics Decoding (OSD). Introduced by Fossorier and Lin \cite{FosLin}, OSD 
represents a novel way to decode binary linear block codes based on the reliability of received symbols and  provides a flexible trade-off between complexity and performance. Among follow-up papers, the work \cite{FosLDPC} studies iterative decoding in conjunction with reliability-based decoding, for medium length LDPC codes. This approach is adapted to the decoding of non-sparse codes in the works \cite{KotTakJinFos} and \cite{JiaNar}, where the authors periodically modify parity check matrices to reduce error propagation. The main idea of our work is to decode LRC codes in two stages, beginning with one step of belief propagation decoding using the local
parities, and continuing with running OSD decoding on the entire code block. We also pursue another, related line of thought, namely, the impact of locality on trellis decoding (trellis complexity) of linear codes. 

Locality constraints introduced in \cite{GopHuaSimYek} were generalized by Kamath et al. \cite{KamPraLalKum} to include stronger local codes and vector alphabets. They were further extended by Sasidharan et al. \cite{hierarchical} who defined a hierarchical structure of 
several levels of locality constraints. At the same time, Wang and Zhang \cite{WanZha} suggested another generalization LRC codes wherein
every coordinate has several independent (disjoint) recovery sets, much like orthogonal parities employed much earlier in threshold
decoding of linear codes \cite{MasseyThesis}. When we wish to emphasize the fact that our codes possess several recovery sets, we
call them LRC codes with availability. In this paper we observe that the decoding algorithm we are having in mind
is fitted well to hierarchical LRC codes and LRC codes with availability, and therefore use them in our examples to quantify the improvements in decoding  performance.

The main class of codes studied in this paper is cyclic LRC codes. These codes were initially studied by Tamo et al. \cite{TBGC2016} and are derived from a more general family of Reed-Solomon-like LRC codes defined earlier in \cite{TamBar}. 
We note that decoding of cyclic codes has been of interest in coding theory due to the rich structure possessed by these codes. Of 
the large number of works devoted to ML decoding of cyclic codes, we mention the paper by Vardy and Be'ery \cite{VarBee} which 
targeted algebraic properties of cyclic codes similar to \cite{TBGC2016}, albeit with no connection to locality constraints. 
The main results in \cite{VarBee} pertain to reducing trellis complexity of cyclic codes by studying 
the direct-sum and concurring-sum structures existing in BCH codes. The direct-sum structure connects their results to LRC codes
with disjoint repair groups, and this is the setting that we pursue in this paper. We note that the constructions in \cite{TamBar} and
\cite{TBGC2016} also form codes with disjoint recovery sets.

\vspace*{.05in}
{\em Our results:} In {\color{black} the first part of} this paper, we suggest a non-iterative, locality-aware decoding scheme for binary cyclic codes with disjoint repair groups and hierarchical locality or availability, built upon the classical OSD \cite{FosLin}. Performance gains in terms of SNR are obtained via a single round of belief propagation based on  local parity checks alone. For codes with disjoint/hierarchical locality, decoding delays are reduced by making use of a simple, quick-look ML decoding stage that relies on the presence of local codes with disjoint supports.

In the second part, which deals with trellis decoding of cyclic codes having composite code lengths, we discuss how to design cyclic codes 
having lower computational complexity through careful introduction of locality. Based on a coordinate ordering that aligns direct-sum 
structures \cite{VarBee}, we obtain an upper bound on state complexity profile of a code with or without locality. Further, the frequency 
domain approach that we adopt, also helps in the better understanding of cyclic codes with $(r,\delta)$-locality and hierarchical locality 
properties.

\section{Locality and Cyclic Codes}\label{sec:prelims}
\subsection{Preliminaries: Codes with Locality}
All the codes discussed in this paper are assumed to be linear of length $n$ and dimension $k$ over a field $\mathbb{F}_q$ unless specified otherwise, and we write their parameters as $[n,k].$ Let $[l]\triangleq\{1,2,\ldots,l\}$ and $[0,l]\triangleq\{0,1,\ldots,l\}$.

 {\color{black}The codes studied in this paper belong to the class of LRC codes \cite{GopHuaSimYek}. We begin with a definition, limiting ourselves to the linear case.}

\begin{definition} \cite{GopHuaSimYek} Let $0\leq i\leq n-1$. The $i^{th}$ coordinate of $\mathcal{C}$ is said to have \emph{locality} $r$ if, for every codeword $\underline{c}\triangleq(c_0,c_1,\ldots,c_{n-1})\in {\mathcal C}$, there exists a subset $\mathcal{I}_i\subset[0,n-1]\setminus\{i\}$ with $|\mathcal{I}_i|\leq r$ such that $c_i$ is a linear combination of the symbols $\{c_j,j\in {\mathcal I}_i\}$. i.e., $c_i=\sum_{j\in\mathcal{I}_i}{\lambda_{i,j}c_j}$, where $\lambda_{i,j}\in\mathbb{F}_q$ and $\lambda_{i,j}\neq 0$ .  Here, both $\mathcal{I}_i$ and $\{\lambda_{i,j}\}$ depend on $i$ but not on $\underline{c}$. A code $\mathcal{C}$ is said to have $r$-all-symbol locality or simply, locality $r$ if, all the coordinates $\{i:0\leq i\leq n-1\}$ have locality $r$. A code having $r$-all-symbol locality is called as an LRC with locality $r$. 
	\end{definition}

{\color{black}The subset ${\mathcal I}_i$ is called the {\em recovery set} of the coordinate $i$. It is clear that every coordinate
$j\in {\mathcal I}_i\cup\{i\}$ can be found from the remaining coordinates in this set, so we call the set ${\mathcal I}_i\cup\{i\}$ a {\em repair group}. Most constructions of LRC codes in the literature, including those in \cite{TamBar,TBGC2016,ChenXiaHaoFu} and the codes considered
in this paper, have the property of disjoint repair groups (for brevity we call them codes with disjoint locality).}

The following generalization of this definition was suggested in \cite{KamPraLalKum}. 
\begin{definition}\cite{KamPraLalKum}
For $0\leq i\leq n-1$, the $i^{\text{th}}$ symbol of a linear code $\mathcal{C}$ over $\mathbb{F}_q$ with generator matrix $\mathbf{G}$ is said to have \emph{$(r,\delta)$-code-symbol locality} if there exists $\mathcal{I}_i\subseteq[0,n-1]$ such that $i\in \mathcal{I}_i$ and $|\mathcal{I}_i|\leq r+\delta-1$, where minimum distance of the punctured code $\mathcal{C}_{\mathcal{I}_i}$ is at least $\delta$. Such a smaller length punctured code will be referred to as a local code. Its corresponding parity checks are termed as local parity checks. 
\end{definition}

 Further, $\mathcal{C}$ is said to possess \emph{$(r,\delta)$-information locality} if there exists an information set $\mathcal{I}\subseteq[0,n-1]$ such that $\mathbf{G}_\mathcal{I}$ ($\mathbf{G}$ punctured to the coordinates in $\mathcal{I}$) has rank $k$ and all $i\in \mathcal{I}$ have  $(r,\delta)$-code-symbol locality. Similarly, if every code-symbol $c_i$, $i \in [0,n-1]$ has $(r,\delta)$-code-symbol locality, the code is said to have $(r,\delta)$-all-symbol locality. Codes with locality $r$ defined above \cite{GopHuaSimYek} form a particular case of this definition for $\delta=2$. For any code $r\le k$, because (unless the distance of $\mathcal C$ is $1$), $\mathbf{G}$ contains $k$ linearly independent 
columns that do not include any chosen coordinate $i$.
 
 \begin{remark} \normalfont We note that the connection between LRC codes and LDPC codes extends to codes having $(r,\delta)$-all-symbol locality. Specifically, for $\delta>2$ the sets of local parities play the same role as local codes in Generalized LDPC codes \cite{len99,bou99}.
This analogy is useful in the derivation of lower bounds on the rate of $(r,\delta)$ LRC codes; see \cite{BargTamoEtAl2015}.
\end{remark}
 
\begin{definition} \cite{WanZha}
A code with $r$-locality is said to have \emph{$t$-availability} if for every $i\in[0,n-1]$, there exist $t$  distinct codewords 
$\underline{c}_1,\underline{c}_2,\dots,\underline{c}_t$ in the dual code $\mathcal{C}^\perp$ such that: 
\begin{itemize}
\item[]
(i)  for any $a \in [t]$, $i \in \text{supp}(\underline{c}_a),$ 
\item[](ii)  for any $a \in [t]$, $|\text{supp}(\underline{c}_a)| \leq r+1$, 
\item[](iii) for any distinct indices $a, b \in [t],$  $\{\text{supp}(\underline{c}_a)\} \cap \{\text{supp}(\underline{c}_b)\} = \{i\}$,
\end{itemize}
where the support of a codeword $\underline{c}\in {\mathcal C}$ is defined as $\text{supp}(\underline{c})\triangleq \{i:c_i\ne 0\}.$
\end{definition} 

We define the support
of a set of codewords $\mathcal{D}$ as $\text{supp}({\mathcal D})\triangleq\cup_{\underline{c}\in {\mathcal D}}\text{supp}(\underline{c}).$

\begin{definition}\cite{hierarchical}
A code is said to possess {\em two-level hierarchical locality} with locality parameters $[(r_\text{mid},\delta_\text{mid}),(r_\text{base},\delta_\text{base})]$, if 
\begin{itemize}
\item[]
(i) for every $i^{\text{th}}$ symbol, there exists a `middle' local code, $\mathcal{C}_i^{\text{mid}}$ with $i\in\text{supp}(\mathcal{C}_i^{\text{mid}})$ such that dim$(\mathcal{C}_i^{\text{mid}})\leq r_\text{mid}$ and $d_{\min}(\mathcal{C}_i^{\text{mid}})\geq \delta_\text{mid}$; 
\item[]
(ii) furthermore, for each $j^{\text{th}}$ symbol of the code $\mathcal{C}_i^\text{mid}$, there exists a `base' local code $\mathcal{C}_{i,j}^{\text{base}}$ with $j\in\text{supp}(\mathcal{C}_{i,j}^{\text{base}})\subseteq\text{supp}(\mathcal{C}_i^{\text{mid}})$ such that dim$(\mathcal{C}_{i,j}^{\text{base}})\leq r_\text{base}$ and $d_{\min}(\mathcal{C}_{i,j}^{\text{base}})\geq \delta_\text{base}$. 
\end{itemize}
\end{definition}

This definition can be extended to more than two levels of local code hierarchy.

\subsection{Cyclic Punctured Codes and Shortened Codes}\label{sec:supportSetDefn}

 Consider an $[n,k]$ cyclic code $\mathcal{C}$ over a field $\mathbb{F}_q$ such that $(n,q)=1$ and $n|(q^m-1)$, where $m$ is the multiplicative order of $q$ modulo $n$. Let $\alpha \in \mathbb{F}_{q^m}$ be a primitive $n^{\text{th}}$ root of unity. Given
 a codeword $\underline{c}=(c_0,c_1,\dots,c_{n-1}),$ let $c(x)=\sum_{i=0}^{n-1}c_ix^i.$ Let $\alpha^{i_1},\alpha^{i_2},\dots$ be the set
 of common zeros of these polynomials in the set of $n^{\text{th}}$ roots of unity. By abuse of terminology, we call
the set $\mathcal{S}(\mathcal{C})\triangleq\{i_1,i_2,\dots\}$ the set of the zeros of the code $\mathcal{C}.$ Recall that 
if $i$ is a zero of $\mathcal{C}$, then so is $qi\text{\,mod\,}n$, {\color{black}and the set of distinct elements among 
$\{(q^{j}i)\text{\,mod\,}n, j=1,2,\dots\}$ is called the cyclotomic coset of $i.$} We note that a cyclic code is fully described by specifying only the representatives of cyclotomic cosets.


Let $\mathcal{I}\subseteq [0,n-1]$ be a subset of code coordinates. The codes  having length $|\mathcal{I}|$ obtained by shortening and puncturing $\mathcal{C}$ to $\mathcal{I}$ are denoted by $\mathcal{C}^\mathcal{I}$ and $\mathcal{C}_\mathcal{I}$, respectively. Let $n_1|n$ and $\nu \triangleq \frac{n}{n_1}$. We define the following set: $\mathcal{I}_i\triangleq\{i,i+\nu,i+2\nu,\ldots,i+n-\nu\}$, where $0\leq i \leq \nu-1$, which may be regarded as being obtained by shifting and then decimating the set $[0, n-1]$. {\color{black}We will call each
such set a {\em length-$n_1$ support set}}. It can be verified that $\mathcal{C}_{\mathcal{I}_i}$'s (and similarly, $\mathcal{C}^{\mathcal{I}_i}$'s) are all identical for $0\leq i\leq \nu-1$ and are moreover cyclic codes of length $n_1$. We refer to these codes as length-$n_1$ punctured (and similarly, shortened) codes of $\mathcal{C}$. The zeros of these codes will be assumed to be computed with respect to the exponents of $\beta=\alpha^\nu$. We cite below a result from \cite{VarBee} which relates the zeros $\mathcal{S}(\mathcal{C}^{\mathcal{I}_i})$ of the shortened code  to $\mathcal{S}(\mathcal{C})$:

\begin{lem} {\emph{\cite{VarBee}}}\label{lem:shortened_code}
	The zeros of the shortened code $\mathcal{C}^{\mathcal{I}_i}$ and the zeros of the code $\mathcal{C}$ are
	related as follows:
	\begin{equation*}
	\mathcal{S}(\mathcal{C}^{\mathcal{I}_i})=\bigg\{{\lambda}\!\!\!\!\!\mod {n_1}: {\lambda} \in \mathcal{S}(\mathcal{C})\bigg\}.
	\end{equation*}	
\end{lem}

We derive the following analogous lemma for punctured codes.

\begin{lem} \label{lem:punc_code}
	The zeros of the punctured code $\mathcal{C}_{\mathcal{I}_i}$ and the zeros of the code $\mathcal{C}$ are
	related as follows:
	          \begin{equation*}
                 	\mathcal{S}(\mathcal{C}_{\mathcal{I}_i})=\bigg\{\lambda\!\!\!\!\!\mod {n_1}:\{\lambda,\lambda+n_1,\ldots,\lambda+(\nu-1)n_1\}\subseteq \mathcal{S}(\mathcal{C}) \bigg\}.	
          \end{equation*}
\end{lem}
The proof of this lemma is given in Appendix \ref{app:punc_lemma_proof}.
{\color{black}We call the set ${\mathcal T}_a^{(n_1)}\triangleq\{a,a+n_1,\ldots,a+(\nu-1)n_1\}$, where $a \in [0, n_1 -1]$, a {\em locality train} corresponding to $n_1.$}

\begin{remark}\label{rem:locality_check}
	\normalfont If $|\mathcal{S}(\mathcal{C}_{\mathcal{I}_i})|>0$, then the code $\mathcal{C}$ has $r$-all-symbol locality, with $r=n_1-|\mathcal{S}(\mathcal{C}_{\mathcal{I}_i})|$.	
\end{remark}

\subsection{Zeros and Locality}\label{sec:zeros_and_loc}

A sufficient condition for a cyclic code to have $r$-all-symbol locality in terms of its zeros was proved in \cite{TBGC2016}
and is stated in the theorem that follows. We use Lemma-\ref{lem:punc_code} to give an alternate proof of this theorem.
\begin{thm} \emph{\cite{TBGC2016}}\label{thm:loc_cyclic_codes_lemma}
	Let $r$ be a positive integer such that $(r+1)|n$. If $\{j,j+(r+1),j+2(r+1),\ldots,j+(\frac n{r+1}-1)(r+1)\} \subseteq\mathcal{S}(\mathcal{C})$ for some $j\in[0,r]$, then $\mathcal{C}$ has $r$-all-symbol locality. 
\end{thm}
\bpf 
Let $n_1=r+1$ and hence, $\nu=\frac{n}{r+1}$. From Lemma-\ref{lem:punc_code}, it follows that $j \in \mathcal{S}(\mathcal{C}_{\mathcal{I}_i})$ for all $0\leq i\leq \nu-1$.  Therefore, any nonzero codeword (polynomial) in each length-$(r+1)$ punctured code $\mathcal{C}_{\mathcal{I}_i}$ has at least two nonzero coefficients. In other words, the distance $d_{\min}(\mathcal{C}_{\mathcal{I}_i})\geq2,$
which implies our claim.
\epf

The above result can be extended to the case of $(r,\delta)$-locality for $\delta>2$, and is given below. This theorem also generalizes a similar result that appears in \cite{ChenXiaHaoFu}.

\begin{thm} \label{thm:r_del_loc_cyclic_codes}
	Let $n_1$ be a positive integer such that $n_1|n$ and $\mathcal{X}\subseteq[0,n_1-1]$. Let $\delta$ be the guaranteed minimum distance for the length-$n_1$ cyclic code having $\mathcal{X}$ and its cyclotomic cosets as zeros. If $\cup_{j\in \mathcal{X}}  \{j,j+n_1,j+2n_1,\ldots,j+n-n_1\} \subseteq \mathcal{S}(\mathcal{C})$, then $\mathcal{C}$ has $(r,\delta)$-all-symbol locality with $r=n_1-\delta+1$.  
\end{thm}

\begin{cor}\emph{\cite{ChenXiaHaoFu}}
	Let $n_1|n$, $\mathcal{X}=\{i_1,i_2,\ldots,i_{\delta-1}\}$ and $d$ be an integer satisfying $(n_1,d)=1$, where $0\leq i_1\leq i_2 \leq \ldots\leq i_{\delta-1}\leq n_1-1$, $i_l=i_1+d(l-1)$, $1\leq l\leq \delta-1$. If $\,\cup_{j\in \mathcal{X}}  \{j,j+n_1,j+2n_1,\ldots,j+n-n_1\} \subseteq \mathcal{S}(\mathcal{C})$, then $\mathcal{C}$ has $(r,\delta)$-all-symbol locality with $r=n_1-\delta+1$.
\end{cor}
\bpf
Follows from application of the BCH bound.
\epf

We invoke Lemma-\ref{lem:punc_code} to show the existence of hierarchical locality in the example below.

\begin{example} 
	\label{eg:cyclic_hier_code}
	\normalfont
	Consider the $[63,33]$ binary cyclic code $\mathcal C$ having zeros $\{0,1,3,5,7,21,27\}$ along with the elements of their  respective $2$-cyclotomic cosets. On account of Lemma \ref{lem:punc_code}, there are three local codes (punctured codes) of length $21$ and dimension $15$ whose zeros are $\{0,3,6,7,12,14\}$ modulo $n_1=21$. Furthermore, each length-$21$ punctured code is a disjoint union of three length-$7$ punctured codes, each of which has $\{0\}$ as its set of zeros. From this, it follows that the hierarchical locality parameters of the code are given by $r_\text{mid}=15$, $\delta_\text{mid}\geq 3$, $r_\text{base}=6$, $\delta_\text{base}= 2$.
\end{example}

\section{A Locality-aware Modification of the Ordered Statistics Decoding (OSD) Algorithm}\label{OrderedStatisticsWithBP}
\subsection{Locality-Aware Modification of the OSD Algorithm} 

We summarize the OSD algorithm due to Fossorier et al. \cite{FosLin} as follows. We consider the channel with additive white Gaussian noise (AWGN), where the transmitted vector is ${\underline{x}}=(x_0\ x_1\ \ldots\ x_{n-1})$, with $x_i \in \{+1,-1\}$. The received vector is ${\underline{y}}={\underline{x}}+{\underline{n}}$, with $n_i \sim\mathcal{N}(0,\frac{N_0}{2}),$ and $N_0$ is the power
spectral density of the noise.  Under the OSD algorithm, the received vector is first ordered in decreasing order of the $|y_i|$'s, which are regarded as reliability values. We call the first $k$ bits in this order which are linearly independent as Most Reliable Independent (MRI) symbols. 
We clip their values to the nearest value of $\{1,-1\}$ and regard them as $k$ information bits of the codeword. The corresponding codeword in $\mathcal{C}$, is selected as the order-$0$ OSD algorithm decoded codeword. This may be followed by an order-$l$ reprocessing stage, where, for $0 \leq i \leq l$, all possibilities of $i$ bits from the $k$ MRI bits will be flipped to obtain $M = \sum_{i=0}^{l} {k \choose i}$ information sets. The $M$ information sets are mapped to their respective codewords and the codeword with least Euclidean distance from $\underline{y}$ will be declared as the order-$l$ OSD output.

In this section, we primarily consider binary linear codes having either (a) disjoint or (b) hierarchical locality. In the case of disjoint locality, the code is assumed to have $r$-all-symbol-locality, with disjoint supports for local (single parity check) codes. For codes with $h$-level hierarchical locality (say, $h=2$) local codes within each level are assumed to have disjoint supports. Furthermore, the base code parity checks here are assumed to include single-parity checks.
\begin{figure}[ht]
	\centering
	\captionsetup{justification=centering}
	\includegraphics[scale=0.4]{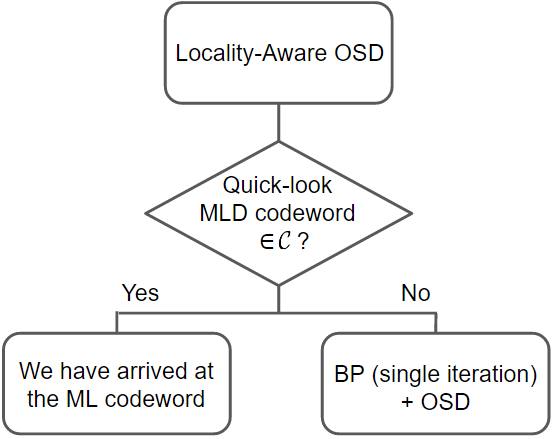}
	\caption{An illustration of the Locality-Aware OSD algorithm.}
	\label{fig:loc_aware_OSD}
\end{figure}

 In the locality-aware OSD algorithm, we first consider a supercode of $\mathcal{C}$ denoted by $\mathcal{C}_{\text{loc}}$, whose dual code is precisely the space spanned by the local parity checks on $\mathcal{C}$. The $\mathcal{C}_{\text{loc}}$ is a concatenation (direct sum) of single parity-check codes (similarly, middle codes) for the disjoint (similarly, hierarchical) locality scenario. We carry out ML decoding (which we call quick-look ML decoding, Q-MLD) on the received vector, $\underline{y}$ with the underlying code assumed to be $\mathcal{C}_{\text{loc}}$.  The disjoint supports of the local codes permit the local codes to be ML-decoded independently.  
 
\subsubsection{Disjoint recovery sets} In this case, ML decoding of a local code can be carried out via  a simple, 2-step procedure: (i) hard-decision decoding of the individual symbols (ii) if this does not yield a valid codeword in the single parity-check code, flip the least reliable bit to obtain the ML local codeword. A little thought will show that, if the $n$-length output vector obtained by putting together the individually ML-decoded local codewords  results in a codeword in $\mathcal{C}$, the decoding procedure can be halted and we have the ML codeword in $\mathcal{C}$. This simple, Q-MLD trick works really well with medium-length codes, particularly in the moderate-to-high SNR range. Suppose for simplicity, that there are $\frac{n}{n_l}$ single-parity locality checks of equal length $n_l$.  It is straightforward to derive the following lower bound on the probability that Q-MLD will succeed: 

 \begin{equation}
 P_{\text{success}}\geq \biggl(1-{n_l\choose 2}\biggl[1-Q\biggl(-\sqrt{\frac{4}{N_0}}\biggr)\biggr]\biggr)^{\frac{n}{n_l}} 
 \label{eq:union_bd_pc}
 \end{equation}

 For a length-$63$ code having nine length-$7$ single parity checks for locality, $P_{\text{success}}\approxeq 0.82$ (empirical) at SNR of $3.5$dB while the estimate given by \eqref{eq:union_bd_pc}, equals $0.77$. At $4.5$dB, the corresponding numbers are given by $0.94$ and $0.93$, respectively.  Thus at $4.5$dB SNR, $>93\%$ of the time, the stage-$1$ decoding will be stopped, after arriving at the optimal codeword choice. This helps in reducing average complexity and decoding delay, when used in conjunction with a buffer. If however, the put-together $n$-length vector is not a codeword in $\mathcal{C}$, a single round of belief propagation based on the single-parity checks alone is carried out, followed by OS order-$l$ decoding.  

\subsubsection{Hierarchical locality case} In the hierarchical case, we arrive at the ML codeword for $\mathcal{C}_{\text{loc}}$ by 
carrying out some decoding procedure, for instance trellis decoding, of each middle code. It is possible to design high-rate codes with 
hierarchical locality, where the middle codes have low trellis-complexity. If Q-MLD does not return a codeword in $\mathcal{C}$, single 
parity checks belonging to base local codes will be effectively utilized by the alternate approach having belief propagation and OSD. As 
middle codes can be designed to be strong codes, this will be a lower probability event compared to the previous case of Q-MLD based on 
single parity checks.  

{\color{black}To illustrate this reasoning, we studied the performance of the code given in Example~\ref{eg:cyclic_hier_code}. 
Consider the pair of embedded codes ${\mathcal{C}}_2\subset \mathcal{C}_1,$ where $\mathcal{C}_1$ has zeros $\{0,1,3,5,7,21\}$ (and their cyclotomic cosets) and the code ${\mathcal{C}}_2$ in addition to these has zeros $\{27,45,54\}$. Here, $\mathcal{C}_2$ is the code with hierarchical locality given in Example~\ref{eg:cyclic_hier_code}. The code  $\mathcal{C}_1$ has
disjoint locality with $r=6$ owing to the presence of 
zeros at $0,7,14,\ldots,56$. We simulated the success rate of Q-MLD for $\mathcal{C}_1$ at SNR 5dB, relying on disjoint repair groups of length 7 and their respective parity equations, and found it to be $\approxeq 0.9650$. We can further improve this outcome by
performing trellis decoding of the middle codes of the code ${\mathcal C}_2$, then the probability of overall decoding success rises to approximately $0.9994$. The trellis decoding steps using the $([0, \infty], + , \min)$ {\em semiring} (see \cite[E.g.~2.5]{McEl}) add $587$ real calculations ($412$ additions and $175$ comparisons) for each of the three length-21 middle codes. Thus, we have a total of $1761$ real calculations ($1236$ additions and $525$ comparisons).} 


  If Q-MLD fails, we perform a simple, one round iteration of BP on a bipartite graph constructed using the single parity local checks having disjoint supports. For each received coordinate $y_i$, we compute the corresponding log-likelihood ratio (LLR) $\ell_i^{(0)}$ ($=\frac{4y_i}{N_0}$, in this case) and feed it to the BP decoder. Let $\{i_1,i_2,\ldots,i_{L}\}$ be the coordinates belonging to a local parity check. Upon completing one round of the BP decoding, we form an updated vector of LLRs, $\underline{\ell}^{(1)}\triangleq(\ell_{1}^{(1)},\ell_{2}^{(1)},\ldots,\ell_{n}^{(1)})$ obtained using the following rule \cite{HagOffPap}:
   \begin{equation}
  \ell_{i_j}^{(1)}=2\tanh^{-1}\biggr[\prod_{a=1, a\neq j}^{L}\tanh\bigg(\frac{\ell_{i_a}^{(0)}}{2}\bigg)\biggr]+\ell_{i_j}^{(0)}
  \end{equation}
The vector  $\underline{\ell}^{(1)}$ is then submitted to the OSD decoder of order $0$ (or higher).
  
Performing one round of message passing (BP) before OSD is achieved at a small computation cost of $O(n)$ real multiplications and additions along with $O(n)$ $\tanh$ and $\tanh^{-1}$ computations.
The conventional OSD order-$0$ algorithm requires $O(n\log n)$ real comparisons and $O(nk^2)$ binary additions/multiplications. For OSD of order $l\ge 1$, the total cost is $O(nk^l)$ real and $O(nk^{l+1})$ binary additions/multiplications. By certain sufficiency tests given in \cite{FosLin}, complexity of OSD can be further reduced.  

If the code has $t$-availability, for each code-symbol, we will update the log-likelihood ratios based on an iteration of BP on the $t$ associated orthogonal parity checks (APP decoding, \cite{MasseyThesis}). This will incur $O(nt)$ $\tanh$, $\tanh^{-1}$ computations and $O(nrt)$ real multiplications.

\subsection{Performance Evaluation Through Simulations}

In the following, we present simulation results which suggest that locality amongst code symbols can be effectively used to obtain significant gains in SNR with negligible complexity overhead in the BP$+$OS stage. Note that these performance gains 
are in addition to the reductions in average complexity and delays obtained through the use of Q-MLD. We reiterate that the BP$+$OS scheme that we use is non-iterative and non-adaptive in nature, making it easier to implement, and yet it gives significant performance gains (see \cite{KotTakJinFos}, \cite{JiaNar} for a discussion of adaptive and iterative decoding techniques).

\begin{figure}[]
	\centering
	\captionsetup{justification=centering}
	\includegraphics[width=4.5in]{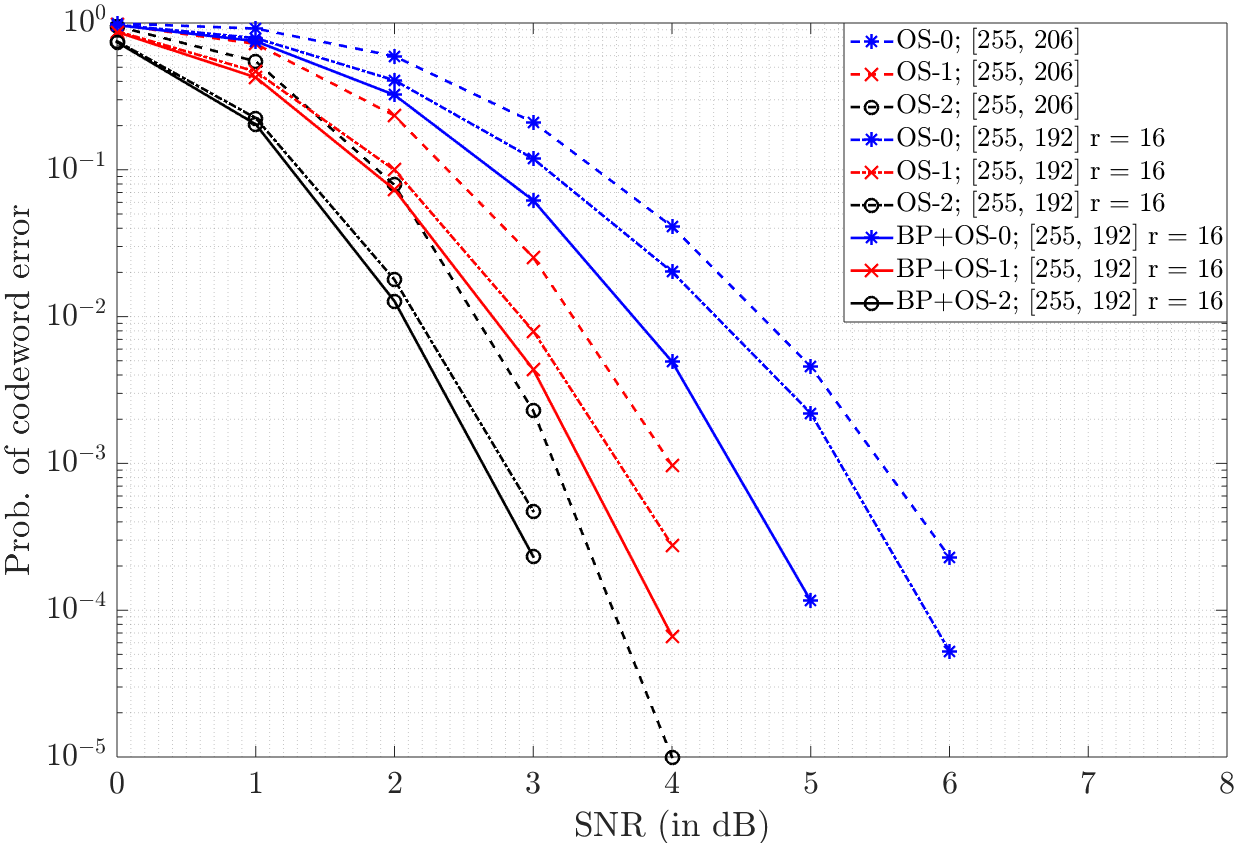}
	\caption{Comparison of performance of a $[255, 206]$ code without locality and a $[255,192]$ code with locality parameter $r = 16$}
	\label{fig:with_without_loc}
\end{figure}

Without loss of generality, we assume transmission of $\underline{1}$ (the all-zero codeword) in the simulations. Fig.~\ref{fig:with_without_loc} presents results relating to two codes; (i) $\mathcal{C}$: the $[255, 206]$ BCH code  having defining zeros at $\{0,1,3,5,7,9,11\}$ and their $2$-cyclotomic cosets (ii) $\mathcal{B}$: a $[255, 192]$ code  with locality, $r = 16$, obtained from $\mathcal{C}$ using  Theorem \ref{thm:loc_cyclic_codes_lemma}. The code $\mathcal{B}$ has defining zeros at $\{0,1,3,5,7,9,11,17,51,85,119\}$ and their $2$-cyclotomic cosets.  To bring out the impact of introducing locality and taking advantage of it, conventional OSD of order-$0$, $1,$ and $2$ is carried out for both codes and in addition, locality-aware OSD with order-$0$, $1$ and $2$ is carried out for $\mathcal{B}$.

\begin{figure}[]
	\centering
	\captionsetup{justification=centering}
	\includegraphics[width=4.5in]{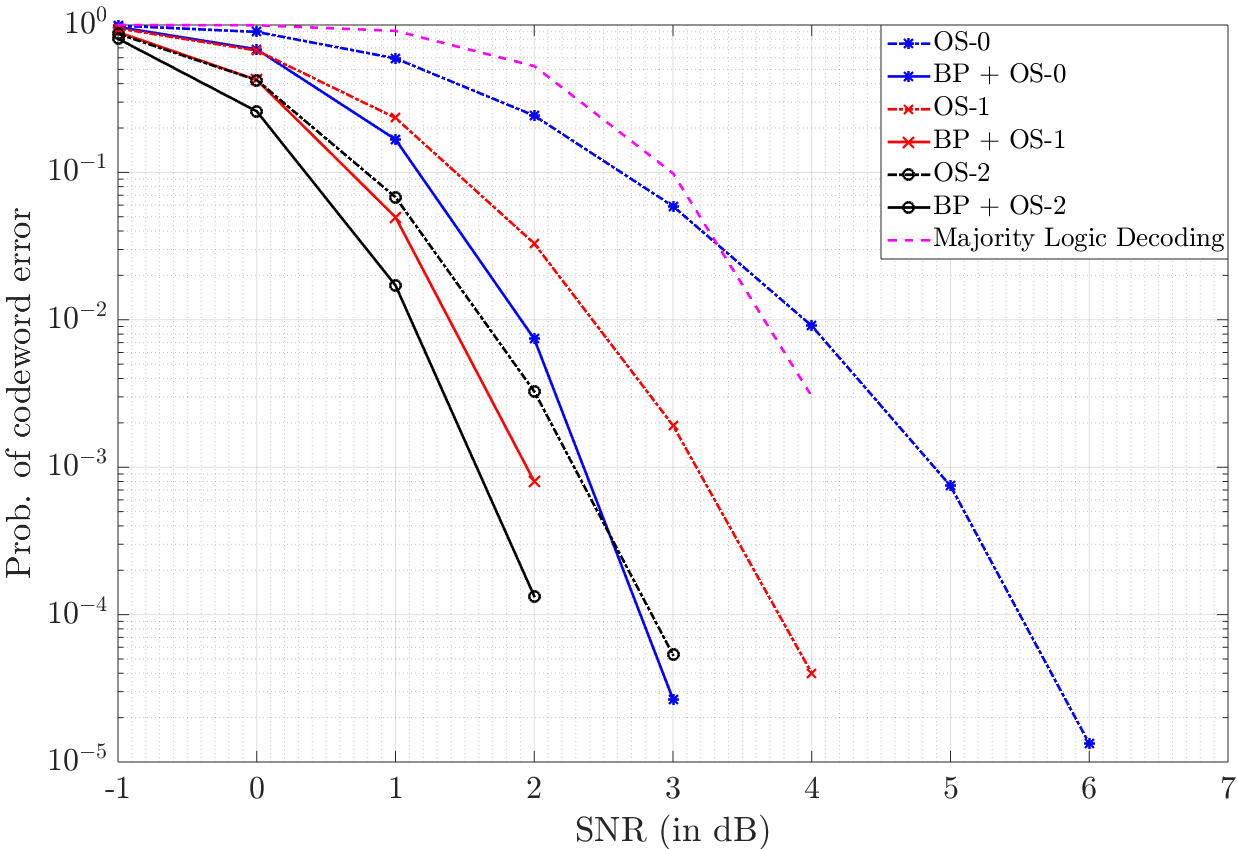}
	\caption{Comparison of OSD, OSD+BP, and majority logic decoding of the type-I DTI code with parameters $[255,175]$, $r=15$, $t = 16$.}
	\label{fig:availability}
\end{figure}

In Fig.~\ref{fig:availability}, we consider the $[255, 175]$ majority logic decodable code \cite[p.290]{LinCos}. This code
has a doubly transitive automorphism group and inherently possesses availability, so we obtain $t = 16$ with $r = 15$. We perform majority logic codeword decoding, conventional OS of order-$0$, $1$ or $2$. In the locality-aware OSD scheme, likelihoods will be obtained for each symbol, based on $t$ orthogonal parity checks, followed by OSD of order-$0$, $1$ and $2$. Here, we observe that BP+order-$0$ OSD outperforms even conventional order-$2$ OSD in the higher SNR range. For both the simulations, similar trends have been observed with respect to Bit-Error-Rate (BER) performance as well, in the same SNR range. For codes having overlapping local parity checks, preliminary experiments are inconclusive and call for further study.

\section{Locality in Trellis Decoding}
In this section, we change the subject and instead of pursuing improved performance of codes in terms of the error probability, 
study the benefits of having locality for reducing {\em decoding complexity}, focusing on trellis decoding.
Trellis representation of a linear block code along with Viterbi decoding enables one to perform efficient
soft-decision ML decoding  \cite{Wolf}, \cite{Forney} , \cite[Ch.~14]{LinCos}.

Let $\underline{c} = (c_0, c_1, \ldots c_{n-1})$ be a codeword of $\mathcal{C}$ and $0\leq i\leq n$. The past shortened code at level-$i$, $\mathcal{P}_{(i)}$ is defined to be the subcode of $\mathcal{C}$ shortened to the first $i$ coordinates, i.e., coordinates $[0,i-1]$. Thus, we have $\mathcal{P}_{(i)} = \{\underline{c} \in \mathcal{C} : c_{i} = c_{i+1} =\ldots =c_{n-1} = 0 \}$. The future shortened code of $\mathcal{C}$ at level-$i$, $\mathcal{F}_{(i)},$ is the subcode of $\mathcal{C}$ shortened to the last $(n-i)$ coordinates $[i,n-1]$ i.e., $\mathcal{F}_{(i)} = \{\underline{c} \in \mathcal{C} : c_{0} = c_{1}= \ldots =c_{i-1} = 0 \}$. Similarly, $\mathcal{P}^{(i)}$ and $\mathcal{F}^{(i)}$ are the codes obtained by puncturing $\mathcal{C}$ to the first $i$ and last $(n-i)$ coordinates, respectively. 
Therefore, we have $\mathcal{P}^{(i)} = \{(c_0, c_1, \ldots, c_{i-1}):\underline{c} \in \mathcal{C}\}$ and $\mathcal{F}^{(i)} = \{(c_{i}, c_{i+1}, \ldots, c_{n-1}):\underline{c} \in \mathcal{C}\}$. The corresponding dimensions of these four codes at level-$i$ are denoted by $p_i$, $f_i$, $p^i$ and $f^i$. By definition we assume that $p_0=p^0=f_n=f^n=0$. 


The trellis diagram is a finite directed graph with a set of vertices and edges $(V, E)$ constructed from the parity-check
matrix of the code \cite{Wolf}, \cite[Ch.9]{LinCos}. We denote the set of vertices at level $i$ of the trellis by $V_i$. 
{\color{black}The {\em state complexity} at level $i$  is defined as $s_i\triangleq\log_q|V_i|,i\in[0,n].$} The following relations hold \cite{Muder}:
\begin{align}
& s_i = k-p_i-f_i=p^i-p_i=p^i+f^i-k=\ f^i-f_i \label{eq:state_complexity}\\
& s_i \leq \min\{k,n-k\} \label{eq:wolf_bound}
\end{align}
The computational complexity of the Viterbi algorithm using trellis decoding is shown to be $2|E|-|V|+1$ \cite{McEl}, where $|E|$ and $|V|$ are the total number of edges and vertices in the trellis, respectively.


Here, we consider cyclic codes having composite code lengths, with $n|(q^m-1)$. The code need not have locality. We assume the natural coordinate ordering that arises from the direct-sum structure \cite{VarBee} existing in these codes. Let $\mathcal{F}_n$ be the set of all factors of $n$ except $1$ and $n$. We will call a subset of factors ${\tau} \triangleq \{x_{1},x_{2},
\ldots, x_{L}\}\subseteq\mathcal{F}_n$ a {\em chain} if $x_{a}|x_{b}$ and $x_a\neq x_b$ for all $a<b$. 

{\color{black}In the remainder of this section, we introduce a coordinate permutation which we will use in our estimates of trellis complexity. We start with a chain $\tau.$}
Using the terminology introduced in Sec.~\ref{sec:supportSetDefn}, for each $l\in[L]$ there exist $\frac{n}{x_l}$ length-$x_l$ support sets. Each of these sets is of the form $\{a,a+x_l,\ldots,a+(\frac{n}{x_l}-1)x_l\}$, for some $a\in[0,x_l-1]$. As explained in 
Sec.~\ref{sec:supportSetDefn}, each of these subsets supports a shortened cyclic code and a punctured cyclic code. {\color{black}Given a chain $\tau$ and the associated support sets, we permute the code's coordinates so that each support set forms a block
of contiguous coordinates. We will denote the resulting ordering of the set $[0,n-1]$ by $\gamma.$ Under this ordering,
each length-$x_l$ support set occupies the coordinates $[(j-1)x_l,jx_l-1]$ for some $j\in\{1,\dots,n/x_l\}$.} We note that the
code obtained by permuting the coordinates of $\mathcal C$ is not necessarily cyclic.

\begin{figure}[ht]
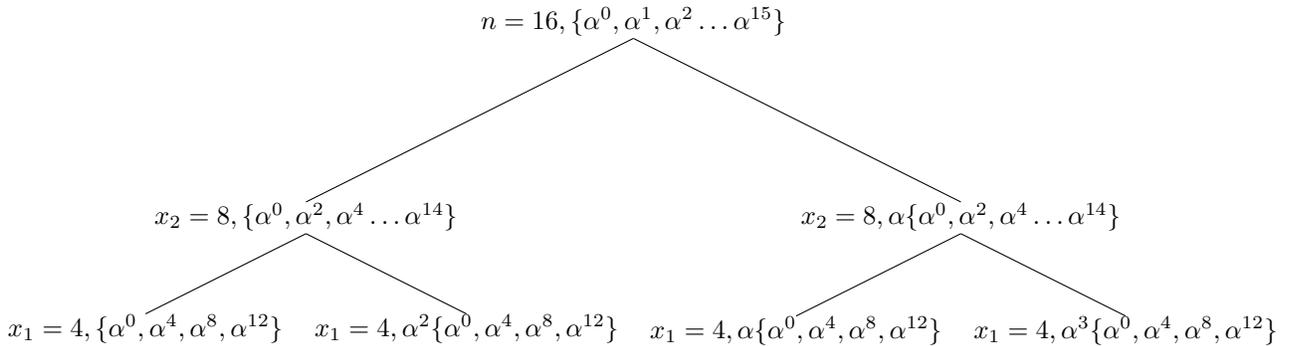

	\centering
	\captionsetup{justification=centering}
	\Tree [ .$n=16,\{\alpha^0,\alpha^1,\alpha^2\ldots\alpha^{15}\}$  [ .$x_2=8,\{\alpha^0,\alpha^2,\alpha^4\ldots\alpha^{14}\}$ [.$x_1=4,\{\alpha^0,\alpha^4,\alpha^8,\alpha^{12}\}$ ] $x_1=4,\alpha^2\{\alpha^0,\alpha^4,\alpha^8,\alpha^{12}\}$ ] [ .$x_2=8,\alpha\{\alpha^0,\alpha^2,\alpha^4\ldots\alpha^{14}\}$ [ .$x_1=4,\alpha\{\alpha^0,\alpha^4,\alpha^8,\alpha^{12}\}$ ] $x_1=4,\alpha^3\{\alpha^0,\alpha^4,\alpha^8,\alpha^{12}\}$ ] ]
	\caption{For a length-$16$ cyclic code over $\mathbb{F}_{17}$ and ${\tau}=\{4,8\}$, the ordering ${\gamma}=(0,4,8,12,2,\ldots,7,11,15)$.}
	\label{fig:permute_tree_explicit}
	
\end{figure}

{\color{black}The coordinate ordering} ${\gamma}$ can be defined with respect to a tree of multiplicative subgroups of the group $\mathbb{F}_{q^m}^\ast$ generated by each of the following field elements: $\alpha^\frac{n}{x_1},\alpha^\frac{n}{x_2},\ldots,\alpha^\frac{n}{x_L}$ and their cosets {\color{black}(here $\alpha$ is a primitive element of $\mathbb{F}_{q^m}$). We use the elements of
the field as locators of the code's coordinates, labeling $i\in[0,n-1]$ as $\alpha^i.$ 
The easiest way to explain the ordering $\gamma$ is by example.
Consider a cyclic code of length $n=16$ over the field $\mathbb{F}_{17}$ and consider the chain ${\tau} = \{4,8\}$.
Let $\alpha$ be a primitive element of $\mathbb{F}_{17}$. Further, let $H_1=\mathbb{F}_{17}^\ast=\langle \alpha\rangle,$ 
$H_2=\langle \alpha^2\rangle,$ and $H_3=\langle \alpha^4\rangle$ be the entire multiplicative group of the field and its subgroups associated with the chain $\tau,$
where $\langle\eta\rangle$ denotes the multiplicative group generated by $\eta$. In Fig.~\ref{fig:permute_tree_explicit} we show
a tree of subgroups and their cosets. The leaves of this tree taken in the natural order (left-to-right according to the figure)
define the coordinate ordering $\gamma$ which is explicitly given as ${\gamma} = 
(0,4,8,12,2,6,10,14,1,5,9,13,3,7,11,15)$.}

\begin{figure}[ht]
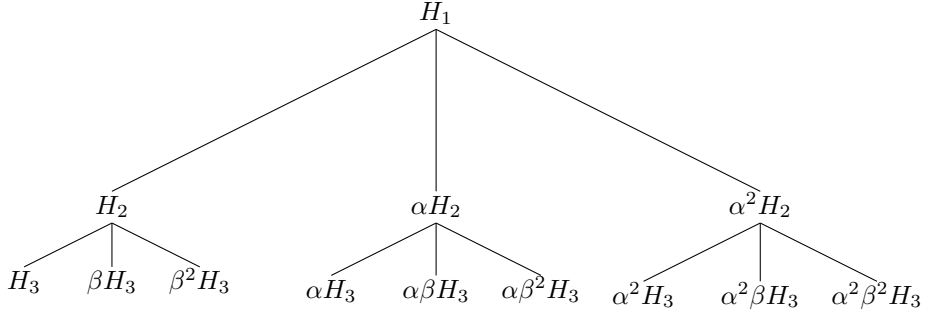

	\centering
	\captionsetup{justification=centering}
	\Tree [.$H_1$ [.$H_2$ [.$H_3$ ] $\beta$$H_3$ $\beta^2$$H_3$ ] [.$\alpha$$H_2$ [.$\alpha$$H_3$ ] $\alpha\beta$$H_3$ $\alpha\beta^2$$H_3$ ]  [.$\alpha^{2}H_2$ [.$\alpha^{2}H_3$ ] $\alpha^2\beta$$H_3$ $\alpha^2\beta^2$$H_3$ ] ]
	\caption{Case $n=63$, ${\tau}=\{7,21\}$. Each leaf gives $7$ coordinates sorted by increasing powers of $\alpha$. For instance, let $\beta H_3=\{\alpha^{i_1},\alpha^{i_2},\ldots,\alpha^{i_7}\}$, with $0\leq i_1<i_2<\ldots<i_7<n$. In this case, the leaf $\beta H_3$ gives the (ordered) coordinates $i_1,i_2,\ldots,i_7$. The ${\gamma}$-ordering corresponds to the natural order of the leaves from left to right. }\label{fig:permute_tree}
	
\end{figure}

To give another example, let $n=63$, ${\tau}=\{7,21\}$, $\beta=\alpha^3$, $\gamma=\beta^3$, $H_1=\langle\alpha\rangle$, and $H_2=\langle\beta\rangle$, $H_3=\langle\gamma\rangle.$ The corresponding ordering ${\gamma}$ is illustrated in Fig.~\ref{fig:permute_tree}. 

If the code $\mathcal{C}$ has the locality property, then using the results in Section \ref{sec:zeros_and_loc}, we observe that this ordering essentially brings the disjoint repair groups to $x_l$ contiguous positions. In the following, by abuse of notation we do not differentiate between
the code ${\mathcal C}$ and its permuted (equivalent) version. 


\subsection{Using locality to reduce state complexity of cyclic codes}

Given an $[n,k]$ cyclic code $\mathcal{C}$ over $\mathbb{F}_q$, it is possible to employ locality to construct a subcode 
$\mathcal{C}_{\text{new}}\subset \mathcal{C}$ with low trellis complexity.
Apart from the obvious dependency of state complexity on the dimension of the punctured code dimensions (viz.~\eqref{eq:state_complexity}), 
the approach of this section is further motivated by empirical analyses suggesting locality to be a ubiquitous feature in cyclic codes of composite  length that have low trellis complexity, with respect to the chain-based orderings we consider. 
For instance, we considered (via a computer search) all binary cyclic codes of length $n = 63$ having low trellis complexity 
(edge count $\leq 50000$,  after fixing the chain to be ${\tau}=\{7,21\}$). Further, we restricted the search to moderate 
code-rates, $0.2 \leq \frac{k}{n}\leq 0.8$. This was because, at very high and very low rates, complexity is bound to be low due to \eqref{eq:wolf_bound}. Interestingly, the computer search revealed that all the codes being considered have $r$-all-symbol locality, with $r\leq 19$ (the codes of dimension $k\leq19$ trivially satisfy this condition), and on an average, the codes had a locality of $10$.

Let $n_j|n$. From Lemma-\ref{lem:punc_code}, if the set of zeros of the code $\mathcal C$ contains a locality train 
$\mathcal{T}_b^{(n_j)} = \{b,b+n_{j},b+2n_{j},\ldots,b+n-n_{j}\}$ for some $b\in[0, n_j-1],$  then $\mathcal C$ has
the locality property, with length-$n_j$ punctured codes as local codes. Our idea in the remainder of this section is to show
that adding locality trains to the set of zeros of the code can be used to reduce the state complexity of resulting subcode.
Let $b\in[0,n_j-1]$ and let $\mathbb{C}_b^{n_j}$ be the $q$-cyclotomic coset modulo $n_j$ that contains $b.$ 
Adding $\mathcal{T}_b^{(n_j)}$ to ${\mathcal S}(\mathcal{C})$ amounts to taking the subcode ${\mathcal C}_{\text{new}}\subset\mathcal C$ whose 
set of zeros is
  $$
    {\mathcal S}(\mathcal{C}_{\text{new}})=\bigcup_{h\in\mathbb{C}_{b}^{n_{j}}}\mathcal{T}_h^{(n_j)}.
  $$

\begin{example}\label{eg:intersection_counting}
	\normalfont
Consider a binary cyclic code $\mathcal{C}$ of length $n=63$ with zeros $\{1,2,4,8,16,32,3,6,12,24,48,33\}$.
We add to $\mathcal{S}(\mathcal{C})$ the following locality trains corresponding to $21$: $T_1=\{0,21,42\}$, $T_2=\{3,24,45\}$. Note that 
adding the sets $T_1,T_2$ is equivalent to adding the zeros $\{0,21,45\}$ and their $2$-cyclotomic cosets. Once these cosets are added to
$\mathcal{S}(\mathcal{C})$, it will also contain the following locality trains:  $T_3=\{6,27,48\}$ and $T_4=\{12,33,54\}$. 
Note that $T_2$, $T_3$ and $T_4$ intersect $\mathcal{S} (\mathcal{C})$ on the subsets $\{3,24\}$, $\{6,27\}$ and $\{12,33\}$, respectively. 
Let $m_{x}$ be the number of locality trains that correspond to 
$x$ and have nonempty intersection with $\mathcal{S}(\mathcal{C})$. In this example, $T_1$ is disjoint from $\mathcal{S}(\mathcal{C})$. Hence, $m_{21}=3.$ 
	\end{example}
	
In the following theorem we quantify our idea of reducing complexity by appending locality trains to the zeros of the code.

\begin{proposition}\label{thm:state_complexity_locality}
Let $\mathcal{C}$ be a cyclic code of composite length and let ${\tau}=\{x_{1},x_{2},\ldots,x_{L}\}$  be a chain. 
Let $\mathcal{C}_{\text{\rm new}}\subset \mathcal{C}$ be a subcode obtained by
adding to ${\mathcal S}(\mathcal{C})$ the locality trains corresponding to $x\in\tau.$ 
Suppose that the coordinates of $\mathcal C$ (and $\mathcal{C}_{\text{\rm new}}$) are permuted according to the ordering $\gamma.$
Then, we have:
	\begin{equation}
	s_{{x}}(\mathcal{C})-s_{{x}}(\mathcal{C}_{\text{\rm new}})=s_{n-{x}}(\mathcal{C})-s_{n-{x}}(\mathcal{C}_{\text{\rm new}})=m_{{x}}\label{eq:state_complexity_diff},
	\end{equation}
where $s_i(\cdot)$ is the state complexity at $i^{th}$ level for the code and
	$m_{x}$ is the number of locality trains added corresponding to $x$, which intersect with $\mathcal{S}(\mathcal{C}).$ 
		\end{proposition}
\bpf
Fix an $x\in\tau$. 
Consider the code $\mathcal{C}_{\text{new}}\subset{\mathcal C}$ obtained by adding some locality trains that corresponding to $x$.
From \eqref{eq:state_complexity} we have
$s_{{x}}(\mathcal{C})-s_{{x}}(\mathcal{C}_{\text{new}})= [ p^{{x}}(\mathcal{C})-p_{{x}}(\mathcal{C})] -[p^{{x}}(\mathcal{C}_{\text{new}})-p_{{x}}(\mathcal{C}_{\text{new}}) ]$. 
By definition of ${\gamma}$   
the first $x$ (and similarly, the last $x$) coordinates of the permuted codewords correspond to length-$x$ support sets. 
This implies that after the permutation, the code obtained by puncturing $\mathcal{C}$ to the first (resp., the last) $x$ coordinates is 
equivalent to a punctured subcode of $\mathcal{C}$ of length $x.$
A similar statement holds for shortened codes as well. It follows that $p^{x}(\mathcal{C})$ and $p^{x}(\mathcal{C}_{\text{new}})$ are 
dimensions of length-${x}$ punctured codes and similarly, $p_{x}
(\mathcal{C})$ and $p_{x}(\mathcal{C}_{\text{new}})$ are the dimensions of length-${x}$ shortened codes. The zero sets for length-$x$ punctured 
and shortened codes are given in Lemma-\ref{lem:punc_code} and Lemma-\ref{lem:shortened_code}. 
As noted above, the union of locality 
trains that correspond to ${x}$ (together with their cyclotomic cosets) results in a disjoint union of 
locality trains. In other words,  every zero in $\mathcal{S}(\mathcal{C}_\text{new})\backslash\mathcal{S}(\mathcal{C})$
is an element in some locality train that corresponds to $x$.   

Every locality train that corresponds to $x$ and is disjoint from $\mathcal{S}(\mathcal{C})$, 
contributes a new zero for the length-$x$ shortened and punctured codes of $\mathcal{C}_\text{new}$ that is not a zero of the corresponding  
length-$x$ shortened and punctured codes of $\mathcal{C}$. This follows from Lemma-\ref{lem:punc_code} and Lemma-\ref{lem:shortened_code}. It is 
clear that length-$x$ punctured (and similarly, shortened) codes of $\mathcal{C}_\text{new}$ are subcodes of length-$x$ punctured (and similarly, 
shortened) codes of $\mathcal{C}$. As state complexity at level $x$ is the difference of the dimensions of length-$x$ shortened and punctured 
code, the locality trains that do not intersect 
$\mathcal{S}(\mathcal{C}),$ will not contribute towards the difference in the state complexity at level $x$ 
(respectively, $(n-x)$).

On the other hand, every added locality train corresponding to $x$ that intersects $\mathcal{S}(\mathcal{C})$, implies the presence of a new 
zero for length-${x}$ punctured code of $\mathcal{C}_{\text{new}}$, whereas for the length-$x$ shortened code of $\mathcal{C}_{\text{new}}$ it 
does not make any difference. This again follows from Lemma-\ref{lem:punc_code} and Lemma-\ref{lem:shortened_code}. Hence, every locality train 
corresponding to $x$ that intersect with $\mathcal{S}(\mathcal{C})$ results in the increase of the difference in the state complexity 
at level $x$ (or $(n-x)$) by $1$, and the proof follows.
\epf

Next we derive a bound on the state complexity of cyclic codes. We begin with explaining the idea.
Let $\mathcal{C}$ be a cyclic code of composite length (with or without locality). We estimate the state complexity of the permuted code
(suppressing the difference between the two versions of the code before and after the permutation).
Suppose that the coordinates of the code ${\mathcal C}$ are permuted with respect to some chain $\tau=\{x_1,\dots,x_L\}$.
It is possible to bound above the state complexity $s_i$ relying on its set of zeros $\mathcal{S}(\mathcal{C}).$ 
Indeed, consider the sections of the trellis at the indices in the set $\tau.$ 
The corresponding values of the state complexity are governed by the dimensions of the respective past and future codes.
From Lemmas~\ref{lem:shortened_code} and -\ref{lem:punc_code}, 
the dimensions of the past codes $\{p^{x_{1}},p^{x_{2}},\ldots,p^{x_{L}}\}$ and $\{p_{x_{1}},p_{x_{2}},\ldots,p_{x_{L}}\}$ can be obtained. Past punctured code dimensions can be expressed in terms of future shortened code dimensions using the rank-nullity theorem, namely
   $$
   p^{n-j}=k-f_{n-j}, \quad j\in[0,n].
  $$
Furthermore, $f_{n-x}=p_{x}$  and $f^{n-x}=p^{x}$, for $x\in\tau$. This follows because the last $x$ 
coordinates form a length-$x$ support set. Hence the code obtained by puncturing (and similarly, shortening) to the last $x$ coordinates 
is permutation-equivalent to length-$x$ punctured (resp., shortened) codes of $\mathcal{C}$. Therefore, we can also find the dimensions of the 
shortened codes $\{p^{n-x_{1}},p^{n-x_{2}},\ldots,p^{n-x_{L}}\}$. 

It is straightforward to see that $\{p^i: i\in[0,n]\}$ is a non-decreasing sequence bounded above by $k$, and the maximum
value of the increase within the $x$-support set is $p^{x}$. Also, $p^{0}\triangleq0$ and  $p^{i+1}-p^{i}\in \{0,1\}$. 
This enables us to construct a sequence $\{\phi_i\}_{i=0}^{n}$ which forms
an upper bound for $\{p^i\}_{i=0}^n$. 
We begin with defining $\phi_i$ for the known instances of $i$:
   $$
     \phi_i=p^i, \quad i\in\{0,x_1,\ldots,x_L,n-x_L,\ldots,n-x_1,n\}
     $$
To fill in the gaps in the sequence, $\{\phi_i\}$ we use the constraints on $\{p^i\}$ mentioned above. Namely, starting with $i=0$, we
either put $\phi_{i+1}=\phi_i$ or $\phi_{i+1}=\phi_i+1,$ as appropriate. Clearly we have
  $$
  p^i\le \phi_i \text{ and }f^i\le \phi_{n-i}
  $$
(the second inequality implied by symmetry). Now let $\mu_i=\phi_i+\phi_{n-i}-k$ and observe from \eqref{eq:state_complexity} that $s_i\le\mu_i.$

Similar relations can be obtained for the dual code using the well-known relation between $\mathcal{S}(\mathcal{C})$ and $\mathcal{S}(\mathcal{C}^\perp)$ ($\beta\in \mathcal{S}(\mathcal{C})$ iff $\beta^{-1}\not\in \mathcal{S}(\mathcal{C}^\perp)$). Using this
together with the fact that the state complexity of $\mathcal C$ and ${\mathcal C}^\perp$ is the same \cite{Forney}, we obtain the bound
   $$
   s_i\leq \min\{\mu_i,\mu_i^\perp\}
   $$
with $\mu_i^\perp$ defined accordingly. Note that this bound is tight for  $i\in \{0, x_1,x_2,\ldots,x_L,n-x_L,n-x_{L-1},\ldots,n-x_1,n\}$.

We can make this argument more explicit by stating an upper bound on the maximum state complexity of the code.

\begin{thm}\label{thm:state_complexity_ub} Let $\mathcal{C}$ be an $[n,k]$ cyclic code with composite length $n$. 
Assume that the coordinates of $\mathcal{C}$ are $\gamma$-ordered based on a chain ${\tau}=\{x_1,x_2,\ldots,x_L\}$.
	Then
	\begin{eqnarray}\label{eq:state_complexity_ub}
	\max_{i\in[0,n]}s_i \leq \min\bigg\{k,n-k,\sum\limits_{i=1}^{L}{\bigg(\frac{x_{i+1}}{x_i}-1\bigg)p^{x_i}+x_1-k}, \sum\limits_{i=1}^{L}{\bigg(\frac{x_{i+1}}{x_i}-1\bigg)(x_i-p_{x_i})+x_1-(n-k)} \bigg\}\nonumber
	\end{eqnarray}
	where we put $x_{L+1}=n$ by definition.
\end{thm}
The proof of this theorem is given in Appendix-\ref{app:max_state_comp_proof}.

The third term under the $\min\{\}$ 
can be  rewritten as $\sum_{i=0}^Lh_{i}$, where $h_{i}\triangleq\frac{x_{i+1}}{x_i}p^{x_i}-p^{x_{i+1}}$, $i>0$, $h_0\triangleq x_1-p^{x_1}$ and $p^{x_{L+1}}=k$. In other words, $h_i,i>0$ indicates the global parities for the length-$x_{i+1}$ punctured code with respect to smaller length-$x_{i}$ punctured codes.  A simple calculation yields that $n-k=h_0\frac{n}{x_1}+h_1\frac{n}{x_2}+h_2\frac{n}{x_3}+\ldots+h_L$. This suggests that, for a given $n$ and $k$, the sum $\sum_{i=0}^Lh_{i}$ can be minimized by making $h_{i}$'s larger for smaller $i$'s (i.e., more locality). The fourth expression accounts for a similar argument about locality of the dual code.

\begin{figure}[ht]
		\centering
		\captionsetup{justification=centering}
		\includegraphics[width=3.5in]{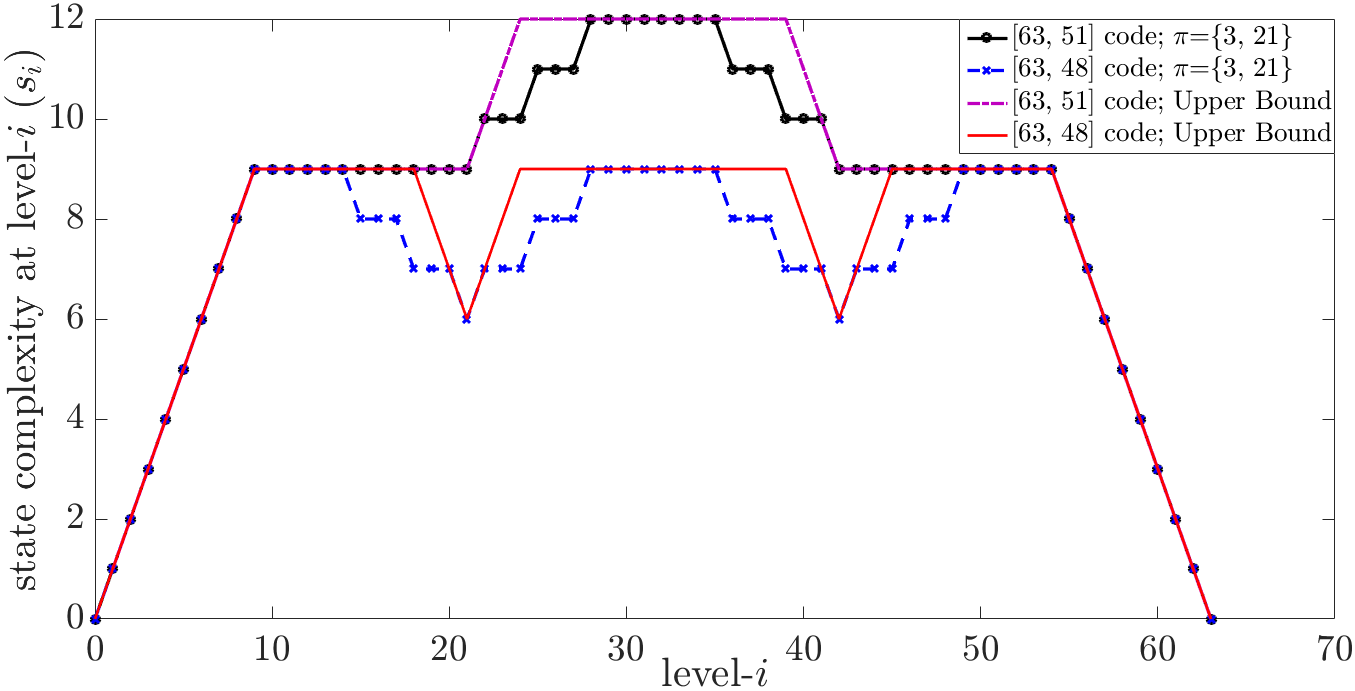}
		
		\caption{State complexity profile and upper bound for codes $\mathcal{C}$ and $\mathcal{C}_1$ of Example \ref{eg:good_code}.}
		\label{fig:trellis_state_complexity}
	\end{figure}

\begin{example} \label{eg:good_code}
	\normalfont
	Let $\mathcal{C}$ be the $[63,51]$ binary BCH code with defining zeros at $\{1,3\}$ and their $2$-cyclotomic cosets (see Table-
	\ref{tab:cyc_cos_63}). We add three locality trains $\mathcal{T}_{3}^{(21)}$, $\mathcal{T}_{6}^{(21)}$ and $\mathcal{T}_{12}^{(21)}$ to $
	\mathcal{S(C)}$, to obtain the $[63, 48]$ BCH-like code $\mathcal{C}_{1}$. All these trains intersect $\mathcal{S}(\mathcal{C})$ and hence, 
	from Proposition-\ref{thm:state_complexity_locality}, $s_{21}(\mathcal{C})-s_{21}(\mathcal{C}_1)=3$. In Table \ref{tab:loc_intro_bch_C1} and 
	Fig. \ref{fig:trellis_state_complexity}, we consider the ${\gamma}$-ordering based on ${\tau} = \{3,21\}$ (which gives the least 
	computational complexity for $\mathcal{C}$ and $\mathcal{C}_1$ among all possible ${\tau}$'s). In order to stress the need for careful 
	locality addition, we consider adding $\mathbb{C}_{7}^{63}$ to $\mathcal{C}$ to obtain the $(63, 45)$ BCH-like code $\mathcal{C}_2$. There 
	are two locality trains $\mathcal{T}_{7}^{(21)}$ and $\mathcal{T}_{14}^{(21)}$ introduced here, with none of them intersecting $\mathcal{S}
	(\mathcal{C})$. In this case, it turns out that, even under the best chain-based permutation for $\mathcal{C}_2$, approximately $12265$ 
	computations ($846$5 real additions $+$ $3800$ comparisons) per information bit are required to decode $\mathcal{C}_2$. For $\mathcal{C}$, $
	\mathcal{C}_1$ and $\mathcal{C}_2$, Theorem \ref{thm:state_complexity_ub} accurately predicts the maximum state complexity.
\end{example}	
	\begin{table}[t]
		\centering
		\begin{tabular}{ c c c c c c } 	\hline		
			$0$ &  &  &  &  & \\ \hline
			$1$ & $2$ & $4$ & $8$ & $16$ & $32$\\\hline
			$3$ & $6$ & $12$ & $24$ & $48$ & $33$\\\hline
			$5$ & $10$ & $20$ & $40$ & $17$ & $34$\\\hline
			$7$ & $14$ & $28$ & $56$ & $49$ & $35$\\\hline
			$9$ & $18$ & $36$ &  &  & \\\hline
			$11$ & $22$ & $44$ & $25$ & $50$ & $37$\\\hline
			$13$ & $26$ & $52$ & $41$ & $19$ & $38$\\\hline
			$15$ & $30$ & $60$ & $57$ & $51$ & $39$\\\hline
			$21$ & $42$ &  &  &  & \\\hline
			$23$ & $46$ & $29$ & $58$ & $53$ & $43$\\\hline
			$27$ & $54$ & $45$ &  &  & \\\hline
			$31$ & $62$ & $61$ & $59$ & $55$ & $47$	\\\hline		
		\end{tabular}
		\caption{$2$-cyclotomic cosets modulo $63$}
		\label{tab:cyc_cos_63}
	\end{table} 
	\begin{table}[t]
		\centering
		\begin{tabular}{ |c|c|c|c|c|c| } 
			
			\hline
			$(n,k)$&No. of&No. of&Approx. computations per& $s_{21}, s_{42}$&$\max_{i} s_i$\\  
			&edges&vertices&information bit & &\\
			& & &(real additions+comparisons) & & \\		
			\hline \hline
			$(63,51)$ & $122876$ & $65534$ & $2409+1124$ &$9$ &$12$ \\ 
			$(63,48)$ & $29180$  & $15998$ & $608+275$  &$6$ &$9$ \\ 
			\hline
		\end{tabular}
		\caption{Trellis complexities of codes $\mathcal{C}$ and $\mathcal{C}_1$ in Example \ref{eg:good_code} with respect to ${\tau} = \{3,21\}$}
		\label{tab:loc_intro_bch_C1}
	\end{table}

 \appendix
\subsection{Proof of Lemma-\ref{lem:punc_code}} \label{app:punc_lemma_proof}

Let $\underline{c} = (c_0,c_1, \ldots ,c_{n-1}) \in \mathcal{C}$ be a codeword and let $c(x)=\sum_{i=0}^{n-1} c_i x^i$ be
the corresponding polynomial. Recall the relation between $\underline{c}$ and its frequency domain representation $\underline{\hat{c}} = (\hat{c}_0,\hat{c}_1, \ldots, \hat{c}_{n-1})$ \cite[Ch.6]{Blahut2003}: For $\lambda=0,1,\dots,n-1$
   $$
     \hat c_\lambda=c(\alpha^\lambda)=\sum_{t=0}^{n-1} c_t\alpha^{t\lambda} \quad\text{ and }\quad
     c_\lambda=\frac 1{n}\sum_{j=0}^{n-1} \hat c_j\alpha^{-\lambda j}. 
   $$
 Therefore $\lambda\in {\mathcal S}(\mathcal {C})$ iff $\hat c_\lambda=0$ for all $\underline{c}\in  \mathcal {C}.$
  
 Now let $\underline{a} = (a_0,a_1, \ldots, a_{n_1 - 1})$ be the codeword obtained by puncturing $\underline{c}$ to the $n_1$ coordinates $\{0,\nu, 2\nu, \ldots, n -\nu\}$. Let $\beta = \alpha^{\nu}$ be a primitive $n_1^{\text{th}}$ root of unity. The frequency domain representation of $\underline{a}$ is given by:
\bea
\hat{a}_{\lambda} &=& \sum\limits_{t = 0}^{n_1 - 1} a_t \beta^{\lambda t} \nonumber \\
&=& \sum\limits_{t = 0}^{n_1 - 1} c_{\nu t} \alpha^{\nu \lambda t} \nonumber\\
&=& \frac{1}{n} \sum\limits_{t = 0}^{n_1 - 1} \sum\limits_{j = 0}^{n - 1} \hat{c}_j \alpha^{-\nu jt} \alpha^{\nu \lambda t} \nonumber\\
&=& \frac{1}{n}\sum\limits_{j = 0}^{n - 1} \hat{c}_j  \sum\limits_{t = 0}^{n_1 - 1} \alpha^{-\nu t (j - \lambda)}\nonumber \\
&=& \frac{n_1}{n}\sum\limits_{j = 0}^{n - 1} \hat{c}_j \Big[\mathbbm{1}(j = \lambda (\text{mod $n_1$}))\Big], \quad
\lambda=0,1,\dots,n_1-1. \label{eq:punc_code_pf}
\eea
Thus $\lambda \text{\,mod\,$n_1$}\in \mathcal{S}(\mathcal{C}_{\mathcal{I}_i})$ if $\lambda\in\mathcal{S}(\mathcal{C}),$ i.e., 
the elements of $\mathcal{S}(\mathcal{C})$ are of the form $j+ln_1,l\in[0,\nu].$

Now let $\lambda = \tilde\lambda + l n_1$, where $l\in[0, \nu-1]$ and $\tilde\lambda\in \mathcal{S}(\mathcal{C}_{\mathcal{I}_i})$. For $\lambda=0,1,\dots,n_1-1$ we have the following:
\bea
\hat{c}_{\lambda} &=& \sum\limits_{t = 0}^{n - 1} c_t \alpha^{\lambda t} \nonumber \\
&=& \sum\limits_{i = 0}^{\nu - 1}\sum\limits_{j = 0}^{n_1 - 1}  c_{i+j \nu} \alpha^{\lambda (i+j \nu)} \nonumber \\
&=& \sum\limits_{i = 0}^{\nu - 1}\alpha^{i \lambda}  \sum\limits_{j = 0}^{n_1 - 1}  c_{i+j \nu} \beta^{\lambda j } \nonumber \\
&=& \sum\limits_{i = 0}^{\nu - 1}\alpha^{i (\tilde\lambda + l n_1)} \Big[\sum\limits_{j = 0}^{n_1 - 1}  c_{i+j \nu} \beta^{\tilde\lambda j} \Big]\nonumber\\
&=& 0 \label{eq:punc_code_pf_2}
\eea

Taken together, \eqref{eq:punc_code_pf} and \eqref{eq:punc_code_pf_2} imply our claim.

\subsection{Proof of Theorem-\ref{thm:state_complexity_ub}} \label{app:max_state_comp_proof}

The first two terms follow from \eqref{eq:wolf_bound}. The third term is obtained by
 a counting argument involving the total number of length-$x_l$ punctured codes of dimensions $\{p^{x_l}\}$, $l \in [L]$, supplied by $\mathcal{P}^{(i)}$ and $\mathcal{F}^{(i)}$ at any level-$i$. Let $\mathcal{C}_{x_l}$ denote the length-$x_l$ punctured code of dimension $p^{x_l}$. We first note the following fact: $\frac{x_b}{x_a}p^{x_a}\geq p^{x_b}$, $\forall$ $x_a,\ x_b:x_a\leq x_b$. This is true because, for each $\lambda \in \mathcal{S}(\mathcal{C}_{x_a})$, $\{\lambda,\ \lambda+x_a, \lambda+2x_a, \ldots, \lambda+x_b-x_a\} \in \mathcal{S}(\mathcal{C}_{x_b})$, from Lemma-\ref{lem:punc_code}. Hence, $p^{x_b}=x_b-|\mathcal{S}(\mathcal{C}_{x_b})|\leq x_b-\frac{x_b}{x_a}|\mathcal{S}(\mathcal{C}_{x_a})|=\frac{x_b}{x_a}(x_a-|\mathcal{S}(\mathcal{C}_{x_a})|)=\frac{x_b}{x_a}p^{x_a}$. If $x_l\nmid i$ for all $l \in [L]$, after counting the smaller punctured codes that contribute to $p^i$ and $f^i$, it can be seen that $s_i\leq \sum_{i=1}^{L}{(\frac{x_{i+1}}{x_i}-1)p^{x_i}+x_1-k}$. Now suppose that $x_L| i$. We have the following chain of inequalities:
\begin{eqnarray*}
	s_i&\leq& p^i+f^i-k\nonumber\\
	&\leq& \frac{n}{x_L}p^{x_L}-k\nonumber\\
	&=& \bigg(\frac{n}{x_L}-1\bigg)p^{x_L}+p^{x_L}-k\nonumber\\
	&\leq& \bigg(\frac{n}{x_L}-1\bigg)p^{x_L}+\bigg(\frac{x_{L}}{x_{L-1}}\bigg)p^{x_{L-1}}-k\nonumber\\
	&=& \bigg(\frac{n}{x_L}-1\bigg)p^{x_L}+\bigg(\frac{x_{L}}{x_{L-1}}-1\bigg)p^{x_{L-1}}+p^{x_{L-1}}-k\nonumber\\
	&\leq& \bigg(\frac{n}{x_L}-1\bigg)p^{x_L}+\bigg(\frac{x_{L}}{x_{L-1}}-1\bigg)p^{x_{L-1}}+\ldots+\bigg(\frac{x_{2}}{x_1}\bigg)p^{x_{1}}-k\nonumber\\
	&\leq& \sum\limits_{i=1}^{L}{\bigg(\frac{x_{i+1}}{x_i}-1\bigg)p^{x_i}+x_1-k}
\end{eqnarray*}

If $x_l\nmid i$ $\forall l\in\{a,a+1,\ldots,L\}$ and $x_{a-1}|i$,
\begin{eqnarray*}
	s_i&\leq& p^i+f^i-k\nonumber\\
	&\leq& \bigg(\frac{n}{x_L}-1\bigg)p^{x_L}+\bigg(\frac{x_{L}}{x_{L-1}}-1\bigg)p^{x_{L-1}}+\ldots+\bigg(\frac{x_{a}}{x_{a-1}}\bigg)p^{x_{a-1}}-k\nonumber\\
	&=& \bigg(\frac{n}{x_L}-1\bigg)p^{x_L}+\bigg(\frac{x_{L}}{x_{L-1}}-1\bigg)p^{x_{L-1}}+\ldots+\bigg(\frac{x_{a}}{x_{a-1}}-1\bigg)p^{x_{a-1}}+p^{x_{a-1}}-k\nonumber\\
	&\leq& \bigg(\frac{n}{x_L}-1\bigg)p^{x_L}+\ldots+\bigg(\frac{x_{a}}{x_{a-1}}-1\bigg)p^{x_{a-1}}+\bigg(\frac{x_{a-1}}{x_{a-2}}\bigg)p^{x_{a-2}}-k\nonumber\\
	&\leq& \sum\limits_{i=1}^{L}{\bigg(\frac{x_{i+1}}{x_i}-1\bigg)p^{x_i}+x_1-k}
\end{eqnarray*}
As $x_i|x_j$ for all $i,j$ such that $1\leq i\leq j\leq L$, these conditions essentially cover all the cases and the proof follows. 
The fourth term is obtained by repeating the same analysis for the dual code.

\bibliographystyle{IEEEtran}
\bibliography{loc_and_dec_NKM}

\end{document}